\documentclass{ifacconf}

\usepackage{graphicx}      % include this line if your document contains figures
\usepackage{natbib}        % required for bibliography
\usepackage{graphicx}
\usepackage{graphicx}      
\usepackage{color}
\usepackage{epstopdf} 
\usepackage{graphicx}
\usepackage{pmat}
\usepackage{amsfonts,amsmath,amssymb}
\usepackage{multirow}
\usepackage{mathtools}
\usepackage[most]{tcolorbox}
\usepackage{xcolor}
\usepackage{adjustbox}
\usepackage{epstopdf}

\newtheorem{Asumm}{Assumption}
\newtheorem{Rem}{Remark}

\newcommand{\Real}{\mathbb R}

\newcommand{\norm}[1]{\left\Vert#1\right\Vert}
\begin{document}
\begin{frontmatter}

\title{Linear Parameter-Varying Embedding of Nonlinear Models with Reduced Conservativeness}% \thanksref{footnoteinfo}} 

\thanks[footnoteinfo]{This work has received funding from the European Research Council (ERC) under the European Union’s Horizon 2020 research and innovation programme (grant agreement nr. 714663).}

\author[First]{Arash Sadeghzadeh} 
\author[First]{Roland T\'{o}th} 

\address[First]{Control Systems, Eindhoven University of Technology, Eindhoven, The Netherlands (e-mail: a.sadeghzadeh@tue.nl, r.toth@tue.nl).}

\begin{abstract}
	In this paper, a  systematic approach is developed to embed the dynamical description of a nonlinear system into a linear parameter-varying (LPV) system representation. Initially, the nonlinear functions in the model representation are approximated using multivariate polynomial regression. Taking into account the residuals of the approximation as the potential scheduling parameters, a principle component analysis (PCA) is conducted to introduce a limited set of auxiliary scheduling parameters in coping with the trade-off between model accuracy and complexity. In this way, LPV embedding of the nonlinear systems and scheduling variable selection are jointly performed such that a good trade-off between complexity and conservativeness can be found. The developed LPV model depends polynomially on some of the state variables and affinely on the introduced auxiliary scheduling variables, which all together comprise the overall scheduling vector. The methodology is applied to a two-degree of freedom (2-DOf) robotic manipulator in addition to an academic example to reveal the effectiveness of the proposed method and  to show the merits of the presented approach compared with some available results in the literature.  
\end{abstract}


\end{frontmatter}
%===============================================================================
\section{Introduction}
The linear-parameter varying (LPV) framework has exten-sively been used to tackle the controller synthesis problems for nonlinear (NL) and time-varying (TV) systems, recently. This modeling methodology is beneficial to extend the mature synthesis techniques for linear systems to NL/TV systems with widespread practical engineering applications ranging from aerospace to process control \citep{HofWer15}. Basically, by introducing some signals so-called scheduling variables, a linear representation is obtained for a NL/TV system. Due to the model simplicity from the controller synthesis standpoint, there exist a flourishing tendency towards this kind of modeling. However, systematic methods to obtain an LPV model for a nonlinear system are scarce. 

Embedding nonlinear models into the LPV model class and also identification-based approaches are the existing methods in order to attain an LPV model. In this context, the identification methods can be categorized as global and local approaches. The local counterpart is based on interpolation of a set of LTI models with the drawback that the time propagation of the scheduling variables is not considered; thus, it is not suitable for systems whose scheduling variables vary excessively fast; on the other hand, the global counterpart relies on a global identification experiment in which both the scheduling variables and the control input have to be excited  persistently, which is not always applicable in practice \citep{Tot10}. Local and global methodologies of LPV model identification have been reviewed and compared on a high-purity distillation column case study in \citet{BacTotLud14}. Some recommendations according to the choice between the methods for a particular process system at hand are provided. An LPV identification technique based on state-space model interpolation of local estimates with a homogeneous polynomial dependence on the scheduling variables in the multisimplex is presented in \citet{CaiCamSwe11}. The problem of control-oriented identification of LPV systems is investigated in \cite{CznMaz03} where obtaining and validating the model is cast as the linear matrix inequality feasibility problem.  Direct embedding of the nonlinear systems into LPV models is a promising approach to this end. Nonetheless, there exist only a few suitable methods which can systematically address this problem.  In \cite{SchTot18}, an LPV embedding approach utilizing linear fractional representation with a nonlinear feedback block is proposed to obtain a model depending affinely on the scheduling variables. A systematic embedding method is presented in \cite{AbbTotPet14} to achieve an LPV state-space representation in the observable canonical form.  An embedding method based on a suitable partitioning of the state vector and with the aid of an apt pre-compensation feedback is deployed in \cite{ChiFalZap03} for the gain-scheduled model predictive control synthesis. The problem of approximation a multi-model system with a low-order LPV  is considered in \citet{PetLof09}. The problem is posed as a model reduction problem to capture the input-output behavior to obtain a low order LPV-approximation.  Constructing affine LPV representations for a nonlinear model is investigated in \citet{KwiBolWer06}.  A criterion presenting the overbounding  of the true nonlinear model is introduced to facilitate choosing an appropriate model from a set of LPV models. 

In this note, a systematic method is investigated to embed a nonlinear model into an LPV one. The objective is to cast the LPV embedding problem into an optimization problem in which both the scheduling variable selection and the conservatism reduction of the model conversion are performed simultaneously. To this end,  the nonlinear functions in the nonlinear representation are approximated using multivariate polynomial regression to obtain an LPV model depending polynomially on some of the state variables which later on will be considered as the scheduling variables. To cope with the approximation error and in order to enhance the model accuracy, the residuals of the approximation  are considered as the potential scheduling variables. Subsequently, to obtain a model with the least possible number of scheduling variables, a principle component analysis (PCA) is carried out to project the possibly correlated residuals into a lower dimensional space. One of the unique features of the obtained LPV model is that the model depends polynomially on some of the scheduling variables contrary to the existing available methods in which the scheduling variables appear affinely in the final model. This leads to the fact that with a less number of scheduling variables a more accurate LPV model is obtained. The overall scheduling variable vector is composed of some of the state variables and a linear combination of some of the mentioned residuals. Compared to the other existing approaches, scheduling variable selection phase is carried out jointly in conservatism reduction phase of LPV embedding process. Due to the fact that the obtained LPV model depends polynomially on the scheduling variables, it can be readily exploited for the controller synthesis by resorting to the relaxation methods to solve the polynomially parameter-dependent linear matrix inequality (LMI) conditions. 

\section{Problem description}
Consider the following continuous-time nonlinear system  
\begin{subequations} \label{Nonlinear-system}
	\begin{align}
	\dot{x}(t)&=F_1(x(t))+G_1(x(t))u(t)  \\ 
	y(t)&=F_2(x(t))+G_2(x(t))u(t)
	\end{align}
\end{subequations}
where $x(t)\in \mathcal{X} \subset \Real^n$ is the state vector, $u(t)\in\Real^m$ is the exogenous input, $y(t)\in \Real^q$ is the system output. $\mathcal{X}$ is assumed to be a bounded polyhedron with known vertices including the origin.  $F_1(x):~\mathcal{X}\to \Real^n$, $F_2(x):~\mathcal{X} \to \Real^q$, $G_1(x):~\mathcal{X}\to \Real^{n\times m}$, $G_2(x):~ \mathcal{X}\to \Real^{q\times m}$ are static real-valued nonlinear analytic functions of $x$ defined on $\mathcal{X}$, which implies that the first order partial derivatives exist on the admissible values for the state variables.  Let us define 
\begin{align*}
 F(x(t))\coloneqq& \left[\begin{array}{cc} F_1(x(t))^\top & F_2(x(t))^\top \end{array} \right]^\top  \\
 =&
\left[ \begin{array}{cccc}
f_1(x(t)) & f_2(x(t)) \cdots & f_{n+q}(x(t)) \end{array} \right]^\top, 
\end{align*}
\begin{align*}
G(x(t))\coloneqq & \left[ \begin{array}{cc} G_1(x(t))^\top & G_2(x(t))^\top \end{array} \right]^\top \\
=&\left[\begin{array}{cccc}
g_{11}(x(t)) & g_{12}(x(t)) & \cdots  &  g_{1m}(x(t)) \\ 
\vdots       & \vdots       &         &  \vdots \\ 
g_{(n+q)1}(x(t)) & g_{(n+q)2}(x(t)  & \cdots  &  g_{(n+q)m}(x(t))
\end{array} 
\right] 
\end{align*}
\begin{align*}
x(t)\coloneqq& \left[
\begin{array}{cccc}
x_1(t)	& x_2(t) & \cdots & x_n(t)	
\end{array}
\right]^\top, \\
u(t)\coloneqq&\left[
\begin{array}{cccc} u_1(t) & u_2(t) & \cdots & u_m(t)
\end{array} 
\right]^\top,\\
\end{align*}

\begin{Asumm}
	All  functions $f_i(x(t))$, $i=1,\cdots,n+q$ satisfy that $f_i(0)=0$.   
\end{Asumm}
Systems in the form of  (\ref{Nonlinear-system}) commonly referred to as control-affine nonlinear systems, often encountered in many practical applications such as process control and mechatronics systems \citep{NijSch90, HenSeb98}.  

The ultimate goal in this paper is to embed the nonlinear system (\ref{Nonlinear-system}) in a linear parameter-varying representation as follows:
\begin{align} \label{LPV-Model}
\dot{x}(t)=&A(\alpha(t))x(t)+B(\alpha(t))u(t), \nonumber \\
y(t)=&C(\alpha(t))x(t)+D(\alpha(t))u(t),
\end{align}
where $\alpha(t)=\left[\begin{array}{cccc} \alpha_1(t)& \alpha_2(t) & \cdots & \alpha_v(t) \end{array}\right]^\top\in \Real^{n_\alpha}$ is the scheduling variable vector and
\[
\underline{\alpha}_i\leq \alpha_i(t) \leq \overline{\alpha}_i,\quad i=1,\cdots,v
\]
with $\underline{\alpha}_i,\overline{\alpha}_i \in \Real$ which are determined in the procedure. To obtain (\ref{LPV-Model}), one should take into account both complexity and accuracy; there exist a trade-off between these two aspects. The objective here is to obtain an accurate LPV model as possible for a pre-chosen number of scheduling variables. The state space matrices in (\ref{LPV-Model}) are assumed to be polynomially parameter-dependent with respect to the scheduling variables. It is worth mentioning that the LPV models depending polynomially on the scheduling variables encompass the traditional affine parameter-dependent ones having been previously considered in the literature \citep{SatPea13,DaaBerGer08,ApkGahBec95}. Nowadays, many approaches have been developed for both performance analysis and controller synthesis for polynomially parameter-dependent LPV models \citep{OliPer07, SadSCL18,CaiCamOli12,SadIJRNC18,SadIET19}. 

\section{LPV modeling}

The main idea behind the embedding of the nonlinear model in an LPV one in this paper is to first extract a polynomially parameter-dependent approximation of the nonlinear existing functions, and then to take into account the residuals of the approximation  as some auxiliary scheduling variables to enhance the overall approximation; however, the number of the introduced scheduling variables should be as small as possible from the practical application perspective both due to complexity and conservativeness. In what follows, different steps of the LPV model construction process are presented.  The main feature of the proposed approach is that the scheduling parameter selection and LPV embedding conservatism reduction are considered in a unique optimization problem.    

\subsection{Polynomial approximation}
At the first step, all the nonlinear existing functions in (\ref{Nonlinear-system}) are approximated as polynomial functions using multivariate polynomial regression method. In other words, all the functions $f_{i}(x(t))$ and $g_{ij}(x(t))$ are approximated with polynomials of user-chosen degree $p$ in order to obtain $\tilde{f}_1(x(t))$, $\cdots$, $\tilde{f}_{n+q}(x(t))$ and $\tilde{g}_{11}(x(t))$, $\cdots$, $\tilde{g}_{(n+q)m}(x(t))$. Suppose that $l(x(t))$ and $\tilde{l}(x(t))$ denote  generic  rep-resentations for either $f_{i}(x(t))$ or $g_{ij}(x(t))$ and their related approximations, respectively. One can obtain the approximated functions as follows:
\begin{align} \label{ApproxFun}
&\tilde{l}(x(t))= 
\sum_{c_1=1}^{n+1}\sum_{c_2=c_1}^{n+1}\cdots\sum_{c_p=c_{p-1}}^{n+1} \eta_{c_1c_2\cdots c_p}~ \rho_{c_1}(t)\rho_{c_2}(t)\cdots \rho_{c_p}(t)
\end{align}
by minimizing the following cost function 
\begin{equation} \label{Cost-Function}
\sum_{j=0}^{N} \left( l(x(jT))-\tilde{l}(x(jT))\right)^2 +\gamma \norm{\eta}_1,
\end{equation}
where
\begin{align*}
\rho(t)  \coloneqq &\left[ \begin{array}{ccc} \rho_1(t) & \cdots & \rho_{n+1}(t) \end{array}\right]^\top \\
=& \left[ \begin{array}{cccc} 1 & x_1(t) & \cdots & x_n(t) \end{array}\right]^\top \in \Real^{n+1}.
\end{align*}
 Note that the coefficients $\eta_{c_1c_2\cdots c_p}$ are the decision variables obtained by minimizing the aforementioned cost function. $T$ is the sampling period.

The term $\gamma \norm{\eta}_1$ in (\ref{Cost-Function}) is considered to enhance the sparsity which means that the approximated functions depend upon a smaller number of monomials. $\gamma$ is the model complexity parameter chosen by the designer and plays the role of a trade-off between the sparsity and the quality of the model. $\eta$ is the decision vector comprising coefficients $\eta_{c_1c_2\cdots c_p}$ appeared in the approximation (\ref{ApproxFun}). Note that there exist more sophisticated methods for this purpose to deploy such as the method in \cite{CanWakBoy08} where a sequence of weighted $l_1$-minimization problems are solved to promote sparsity.

Subsequently, the related residuals are readily derived as follows for all $i=1,\cdots, n+q$ and $j=1,\cdots,m$.

\begin{align*}
e_i^f(x(t))&=f_i(x(t))-\tilde{f}_i(x(t)),	\\ 
e^g_{ij}(x(t))&=g_{ij}(x(t))-\tilde{g}_{ij}(x(t)),	
 \end{align*}

\subsection{Factorization}
So far, (\ref{Nonlinear-system}) can be written as follows:
\begin{align*}
\dot{x}(t)&=\tilde{F}_1(x(t))+E^f_1(x(t))+\left(\tilde{G}_1(x(t))+E^g_1(x(t))\right)u(t), \\
y(t)&=\tilde{F}_2(x(t))+E_2^f(x(t))+\left(\tilde{G}_2(x(t))+E^g_2(x(t))\right)u(t),
\end{align*}
where
\begin{align*}
\tilde{F}_1(x(t))&=\left[\begin{array}{cccc} \tilde{f}_1(x(t)) & \tilde{f}_2(x(t)) & \cdots & \tilde{f}_n(x(t)) \end{array}\right]^\top, \\
E_1^f(x(t))&=\left[\begin{array}{cccc} e_1^f(x(t)) & e_2^f(x(t)) & \cdots & e_n^f(x(t)) \end{array}\right]^\top,
\end{align*}
\begin{align*}
\tilde{G}_1(x(t))&= \left[\begin{array}{cccc} \tilde{g}_{11}(x(t)) & \tilde{g}_{12}(x(t)) & \cdots & \tilde{g}_{1m}(x(t)) \\ \vdots & \vdots & & \vdots \\ \tilde{g}_{m1}(x(t)) & \tilde{g}_{m2}(x(t)) & \cdots & \tilde{g}_{nm}(x(t))
\end{array}\right], \\
E_1^g(x(t))&=\left[\begin{array}{cccc} e_{11}^g(x(t)) & e_{12}^g(x(t)) & \cdots & e_{1m}^g(x(t)) \\ \vdots & \vdots &  &  \vdots \\ e_{n1}^g(x(t)) & e_{n2}^g(x(t)) & \cdots & e_{nm}^g(x(t))
\end{array}\right],
\end{align*}
and the definitions for $\tilde{F}_2(x(t))$, $E^f_2(x(t))$, $\tilde{G}_2(x(t))$, and $E^g_2(x(t))$  are similar to those of  $\tilde{F}_1(x(t))$, $E^f_1(x(t))$, $\tilde{G}_1(x(t))$, and $E^g_1(x(t))$, respectively, but, due to brevity, are omitted. 

In order to obtain the representation (\ref{LPV-Model}), $E_1^f(x(t))$ and $E_2^f(x(t))$ need to be factorized. This means that they should be decomposed as $E_1^f(x(t))=\bar{E}_1^f(x(t))x(t)$ and $E_2^f(x(t))=\bar{E}_2^f(x(t))x(t)$. Bear in mind that the polynomial approximations $\tilde{F}_1(x(t))$ and $\tilde{F}_2(x(t))$ are easily factorisable since they are constituted from constant terms and polynomial terms in $x(t)$. It is worth mentioning that the polynomial approximations for the functions $f_1(x(t)),\cdots,f_{n+q}(x(t))$ should be at least of degree 1 for this purpose. Note that we have
\begin{align*}
&f_i(x(t))=\tilde{f}_i(x(t))+e_i^f(x(t)) \\
&\quad=\left(\underbrace{\tilde{f}_i(x(t))-\tilde{f}_i(0)}\right)+\left(\underbrace{e_i^f(x(t))+\tilde{f}_i(0)}\right). \\
&\qquad\qquad\quad \bar{f}_i(x(t)) \qquad\qquad\qquad \bar{e}_i^f(x(t))
\end{align*}

 Under mild assumption that the functions $f_i(x(t))$  do not contain singular points for $x(t)\in \mathcal{X}$, by taking advantage of the presented method in \cite{SchTot18},  we can factorize $E_1^f(x(t))$ and $E_2^f(x(t))$. Note that 
 \[\bar{e}_i^{f}(0)=f_i(x(t))|_{x(t)=0}-\tilde{f}_i(x(t))|_{x(t)=0}+\tilde{f}_i(0)=0\]
 when $f_i(0)=0$ which have been previously assumed. This assumption is requisite for the considered factorization. Now, following the method in \cite{SchTot18}, the factorization may be performed as follows:
\begin{equation*}
\bar{e}_i^{f}(x(t))=\left[ \begin{array}{cccc}\bar{e}_{i1}^f(x(t)) &\bar{e}_{i2}^f(x(t)) & \cdots & \bar{e}^f_{in}(x(t))\end{array} \right]x(t),
\end{equation*}
with
\begin{equation*}
\bar{e}_{ik}^f(x(t))=\left\{
\begin{array}{l}
\frac{\bar{e}_i^f(\breve{x}_k(t))-\bar{e}_i^f(\breve{x}_{k-1}(t))}{x_k}\quad \text{if}~ x_k \neq 0, \\
\left. \frac{\partial\bar{e}_i^f(\breve{x}_k(t))}{\partial x_k}\right\vert_{x=\breve{x}_{k-1}} \quad\qquad \text{if}~ x_k=0,
\end{array}
\right.
\end{equation*}
and
\begin{equation*}
\bar{e}_{i1}^f(x(t))=\left\{
\begin{array}{l}
\frac{\bar{e}_i^f(\breve{x}_1(t))}{x_1} ~~\quad\quad\quad\text{if}~ x_1 \neq 0, \\
\left. \frac{\partial\bar{e}_i^f(\breve{x}_1(t))}{\partial x_1}\right\vert_{x=0} \quad \text{if}~ x_1=0,
\end{array}
\right.
\end{equation*}
for $i=1,\cdots,n+q$ and $k=1,\cdots,n$, where
\[
\breve{x}_k\coloneqq \left[\begin{array}{ccccccc} x_1(t) & x_2(t)& \cdots & x_k(t) & 0 & \cdots & 0\end{array}\right]^\top\in \Real^n.
\]

As we stated before, $\tilde{F}_1(x(t))$ and $\tilde{F}_2(x(t))$ after removing the constant terms can also be readily factorized since they are polynomial functions with respect to the state variables and at least one of the state variables appears in each of their constituting terms. Note that, however, the factorization is non-unique, and it should be further investigated which factorization results in better controll-ability index, conservatism reduction, and control-oriented LPV embedding. Now, it is assumed that  $\bar{f}_i(x(t))$  are factorized as follows:
\begin{equation} \label{Facotr_Poly}
\bar{f}_i(x(t))=\left[ \begin{array}{cccc} \beta_{i1}(x(t)) & \beta_{i2}(x(t)) & \cdots &\beta_{in}(x(t)) \end{array}\right] x(t).
\end{equation}

Considering the fact that $f_i(x(t))=\bar{f}_i(x(t))+\bar{e}_i^f(x(t))$, we finally obtain
\begin{equation*}
f_i(x(t))=\left[\begin{array}{cccc} a_{i1}(x(t)) & a_{i2}(x(t)) & \cdots & a_{in}(x(t)) \end{array} \right]x(t),
\end{equation*}
where
\begin{equation} \label{a_ik}
a_{ik}(x(t))\coloneqq \beta_{ik}(x(t))+\bar{e}_{ik}^f(x(t)).
\end{equation}

Additionally, let us define
\begin{equation} \label{b_ij}
b_{ij}(x(t))\coloneqq \tilde{g}_{ij}(x(t))+e^g_{ij}(x(t)).
\end{equation}
To wrap up this section, an LPV model is derived for the nonlinear system (\ref{Nonlinear-system}) as follows:
\begin{align} \label{LPV-Initial}
\dot{x}(t)&=A(x(t),e(t))x(t)+B(x(t),e(t))u(t), \nonumber\\
y(t)&=C(x(t),e(t))x(t)+D(x(t),e(t))u(t),
\end{align}

by introducing the residual signals as the auxiliary scheduling variables. Where 

\begin{align} \label{State-Space-Matrices}
&\begin{pmat}[{|}]
A(x(t),e(t)) & B(x(t),e(t)) \cr\-
C(x(t),e(t)) & D(x(t),e(t)) \cr
\end{pmat}=\\ \nonumber
&\begin{pmat}[{...|..}]
a_{11} & a_{12} & \cdots  & a_{1n} & b_{11} & \cdots & b_{1m} \cr
\vdots & \vdots &         & \vdots & \vdots &        & \vdots \cr
a_{n1} & a_{n2} & \cdots  & a_{nn} & b_{n1} & \cdots & b_{nm} \cr\-
a_{(n+1)1}& a_{(n+1)2} & \cdots & a_{(n+1)n} & b_{(n+1)1} & \cdots & b_{(n+1)m}\cr
\vdots    & \vdots     &        &  \vdots    & \vdots     &        &  \vdots  \cr
a_{(n+q)1}& a_{(n+q)2} & \cdots & a_{(n+q)n} & b_{(n+q)1} & \cdots &   b_{(n+q)m} \cr
\end{pmat},
\end{align}
with
\begin{equation*}
e(t)=\left[ \begin{array}{cc}
e^f(t) & e^g(t)
\end{array} \right]^\top \in \Real^{(n+q)(n+m)\times 1},
\end{equation*}
where
\begin{align*}
&e^f(t)= \\ 
&\left[\begin{array}{cccccc} \bar{e}_{11}^f(x(t)) & \cdots &\bar{e}_{1n}^f(x(t)) &\bar{e}_{21}^f(x(t)) & \cdots  & \bar{e}_{(n+q)n}^f(x(t)) \end{array} \right], 
\end{align*}
\begin{align*}
&e^g(t)= \\
& \left[ \begin{array}{ccccccc}
e_{11}^g(x(t)) & \cdots & e_{1m}^g(x(t)) & e_{21}^g(x(t)) &  \cdots & e_{(n+q)m}^g(x(t)) 
\end{array}\right]. 
\end{align*}
 For the ease of notation, the dependencies of $a_{ik}$ and $b_{ij}$, given by (\ref{a_ik}) and (\ref{b_ij}),  upon $x(t)$ and $e(t)$ are dropped. $a_{ik}$ and $b_{ij}$ are polynomial functions with respect to $x(t)$ and affine functions in the residuals  $e(t)$. 
 
\begin{Rem}
	As it is stated in \citet{SchTot18}, the aforementioned employed factorization is not unique. By changing the order in which the variables $x_i$ are considered, different factorizations of the residuals could be obtained.
\end{Rem}
 
\subsection{PCA-based scheduling variable selection} \label{PCA_section}
Note that the obtained LPV model (\ref{LPV-Initial}) does not have reasonable number of the scheduling variables; indeed, there exist  generally $(n+m)(n+q)$  scheduling variables, a huge number even for a simple nonlinear system. To tackle this drawback, one can resort to principle component analysis (PCA) to reduce the number of the scheduling variables. Inspiring by the presented method in \citet{KwiWer08}, first generate the following data matrix

\begin{equation*}
\Pi=\left[\begin{array}{cccc} e(0)	& e(T) &  \cdots & e(NT) \end{array} \right]
\end{equation*}
 Then, the row $\Pi_i$ of this matrix is normalized by an affine law $\mathcal{N}_i$ to obtain scaled, zero mean data
\[
\Pi^n_i=\mathcal{N}_i(\Pi_i) 
\]
leading to the normalized matrix $\Pi^n=\mathcal{N}(\Pi)$ that can be employed for the PCA. Now, consider the singular value decomposition of $\Pi^n$ as follows 
\begin{equation} \label{SVD}
\Pi^n=\left[ \begin{array}{cc} U_s & U_n \end{array} \right]
\left[ \begin{array}{ccc}
\Sigma_s & 0       & 0 \\
0        &\Sigma_n & 0
\end{array}
\right]
\left[ \begin{array}{c}
V_s' \\
V_n'
\end{array}
\right]
\end{equation}
where $U_s$, $\Sigma_s$, and $V_s$ correspond to $v$ significant singular values. In case the residuals are correlated then the insignificant singular values can be neglected to obtain an approximation for the residuals. In this regard, introducing the following set of scheduling variables  
\[
\theta(t)\coloneqq \left[ \begin{array}{cccc} \theta_1(t) & \theta_2(t) & \cdots & \theta_v(t) \end{array}\right]^\top=U_s^\top\mathcal{N}(e(t))
\]
enables us to obtain an approximation $\tilde{e}(t)$ for the residuals as follows:

\begin{align*}
\tilde{e}(t)\coloneqq&\left[ \begin{array}{cccc} \tilde{e}_1(\theta(t)) & \tilde{e}_2(\theta(t)) & \cdots & \tilde{e}_{(n+q)(n+m)}(\theta(t)) \end{array}
\right] \\
=& \mathcal{N}^{-1}(U_s\theta(t)) \in \Real^{(n+q)(n+m) \times 1}
\end{align*}
using a less number of scheduling variables. Here,  $\mathcal{N}^{-1}$ denotes row-wise rescaling. 

Note that the approximation $\tilde{e}(t)$ is a multivariate affine function with respect to the newly introduced scheduling variable vector $\theta(t)$. As a matter of fact, there exists a trade-off between the number of the scheduling variables $v$ and the desired accuracy of the model. 

As a measure of the quality of the approximation, one can consider the following criterion which is the fraction of total variation $v_m$ and defined as follows:

\begin{equation*}
v_m\coloneqq \frac{\sum_{i=1}^v \sigma_i^2}{\sum_{i=1}^{(n+q)(n+m)} \sigma_i^2}
\end{equation*}
where $\sigma_i$ denote the singular values in (\ref{SVD}). 

It is worth mentioning that we employ PCA strategy to cope with the residuals, then, based on the required accuracy, the scheduling variables are chosen. Actually, the scheduling variables selection and the conservatism reduction are performed jointly.   The  $v_m$ value denotes how much of the residuals are covered by the chosen scheduling variables. 
 
\subsection{LPV model}
To recapitulate, one can readily obtain an LPV model (\ref{LPV-Model}) for the nonlinear system (\ref{Nonlinear-system}) with the state space matrices given by (\ref{State-Space-Matrices}), where

\begin{align*}
a_{ik}(\alpha(t))&=\tilde{e}_{(i-1)n+k}(\theta(t))+\beta_{ik}(x(t)) \\
b_{ij}(\alpha(t))&=\tilde{e}_{n(n+q)+(i-1)n+j}(\theta(t))+\tilde{g}_{ij}(x(t))
\end{align*}
for $i=1,\cdots,n+q$, $k=1,\cdots,n$, and $j=1,\cdots,m$.
Note that the overall  scheduling variable vector $\alpha(t)$ comprises the vector $\theta(t)$,  those state variables that remain in $\beta_{ik}(x(t))$ after the factorization (\ref{Facotr_Poly}), and those state variables that exist in $\tilde{g}_{ij}(x(t))$.  Due to the fact that $\theta(t)$ depends affinely on the  residuals $e(t)$ which are nonlinear functions of $x(t)$, one can readily compute the upper bounds $\overline{\alpha}_i$ and lower bounds $\underline{\alpha}_i$ for all the scheduling variables knowing the related intervals of the state variables.  

\section{Numerical simulation}

In this section, simulation studies are carried out to  demonstrate the superiority of the proposed method in comparison with some available approaches. 
\subsection{Example 1} 
\begin{figure*}[t]
	\begin{center}
		\includegraphics[width=19cm]{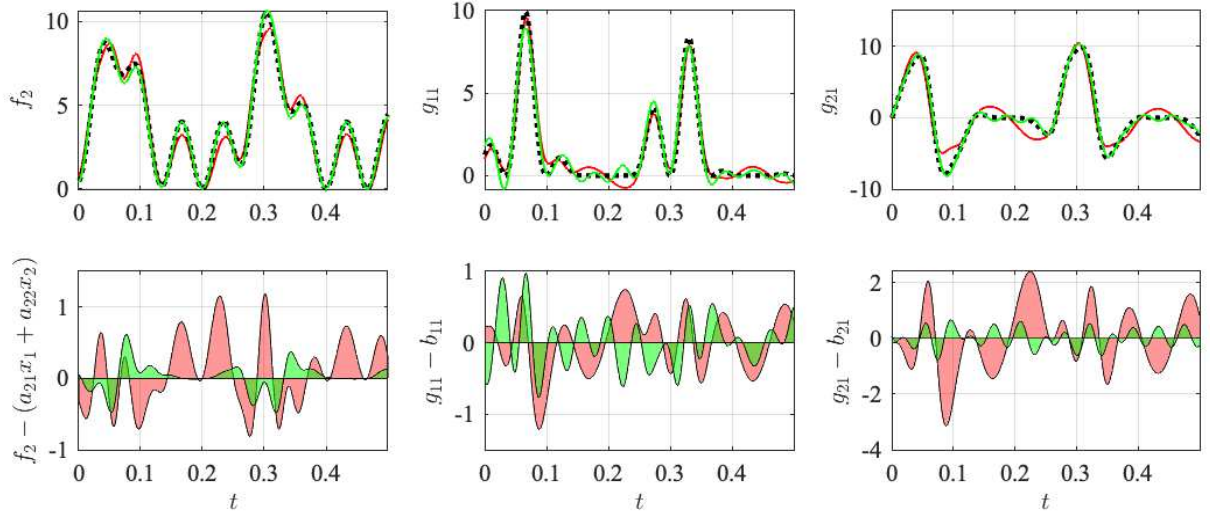}    % The printed column width is 8.4 cm.
		\caption{\raisebox{+1.5ex}{\colorbox{green!50}{}} The related results of the second scenario in Example 1; \raisebox{+1.5ex}{\colorbox{red!40}{}}  The related results of the first scenario in Example 1. First row: the actual values for $f_2(x)$, $g_{11}(x)$, and $g_{21}(x)$ (dotted line) and their LPV representation counterparts for the two scenarios. Second row: the errors between the true and the approximated functions.}
		\label{Exam1-Fig}
	\end{center}
	\hrulefill
	\vspace*{4pt}
\end{figure*}
Consider the nonlinear system (\ref{Nonlinear-system}) with
\begin{align*}
\left[\begin{array}{c} f_1(x) \\ f_2(x) \\f_3(x) \end{array}\right]&=\left[ \begin{array}{c}
5x_2+10x_1x_2-2x_1^3+3x_1x_2\sin(\frac{\pi}{2}x_2)\\
7x_1^4+4x_2^2 \\
x_1
\end{array}
\right] \\
\left[\begin{array}{c} g_{11}(x) \\ g_{21}(x) \\g_{31}(x) \end{array}\right]&=\left[ \begin{array}{c}
10x_1^3\cos(\frac{\pi}{2}x_2)\\
10x_1^2\sin(\frac{\pi}{2}x_2) \\
1
\end{array}
\right]
\end{align*}
We consider two scenarios for this simple nonlinear system. First scenario: 1st order polynomial approximations for $f_i$ functions and zero-order approximation for $g_{ij}$ functions. Second scenario: 3rd order polynomial approximations for all the functions. In these cases, for different numbers of scheduling variables, the LPV models are synthesized. The quality of the approximations which is evaluated by the value of $v_m$ are reported in  Table \ref{Exam1-Tab}. Obviously, increasing the number of the scheduling variables enhance the LPV modeling. It is worth mentioning that for the first scenario all the scheduling variables are those  introduced by PCA-based scheduling variable selection procedure explained in Section \ref{PCA_section}. For the second scenario, due to the fact that the  $f_i$ functions are approximated by 3rd order polynomials, one should always consider $x_1$ and $x_2$ as the scheduling parameters in addition to the introduced scheduling variables in Section \ref{PCA_section} since after factorization $x_1$ and $x_2$ remain in the coefficients $\beta_{ik}(x(t))$. 

The nonlinear functions  $f_2(x)$, $g_{11}(x)$, $g_{21}(x)$ and their counterparts in the LPV representation, namely,  $a_{21}(\alpha)x_1$ $+a_{22}(\alpha)x_2$, $b_{11}(\alpha)$, and $b_{21}(\alpha)$ are depicted in Fig. \ref{Exam1-Fig} for the first and second scenarios using 3 scheduling variables. Moreover, in the second row of the same figure the difference between the actual functions and their counterparts in the  LPV representation are depicted.  For the first scenario the scheduling variables are  $\alpha=(\theta_1,\theta_2,\theta_3)$ and for the other $\alpha=(x_1,x_2,\theta_1)$ are the actual scheduling variables. According to Fig. \ref{Exam1-Fig}, it is obvious that the second scenario leads to better results. Even though the $v_m$ value for the second scenario is less than that of the first scenario in these two cases (see the colored values in Table \ref{Exam1-Tab}), the obtained results by the second scenario are better since the residuals are intrinsically different, and it is obvious that the residuals related to the second scenario are less than that of the first scenario; therefore, even with a less value of $v_m$ better results are obtained.  Bear in mind that $v_m$ for each case reveals that how much of the residuals in that scenario are covered by introduced scheduling variables.   Furthermore, for the quantitative comparison, mean square error (MSE) between the actual values of the nonlinear functions and their counterparts in the LPV modeling are reported in Table \ref{MSE-table} for both scenarios using 3 scheduling variables. 

\begin{table}
	\centering
	\caption{$v_m$ values for the first and second scenarios  for different number of scheduling variables (No. Sch.)}
	\begin{tabular}{cccccc}
		\hline
		No. Sch.                      &  1            &  2           & 3                                                           &  4           &  5            \\
		\hline
		First scenario              &  0.6273   & 0.8835  & \bfseries{\textcolor{red!40}{0.9609}}     & 0.9938   & 0.9998  \\
		Second scenario         & ---        & ---       & \bfseries{\textcolor{green!50}{0.3009}} & 0.5285   & 0.7137        \\
		\hline           
		\label{Exam1-Tab}
	\end{tabular}
\end{table}

\begin{table}
	\centering
	\caption{MSE values for the first and second scenarios where $c^f_i=f_i(x)-(a_{i1}(\alpha)x_1+a_{i2}(\alpha)x_2)$ and $c^g_j=g_{i1}(x)-b_{i1}(\alpha)$ for $i=1,2$. }
	\begin{tabular}{cccccc}
		\hline
		                                   &     $c^f_1$    &    $c^f_2$   &    $c^g_1$        &   $c^g_2$     \\
		\hline
		First scenario              &     0.1244   &  0.2476     &     0.1876       &  1.4796       \\
		Second scenario         &     0.0200   &  0.0369     &    0.1237        & 0.1021        \\
		\hline           
		\label{MSE-table}
	\end{tabular}
\end{table}

\subsection{Example 2}
\begin{figure*}[t]
	\begin{center}
		\includegraphics[width=18cm]{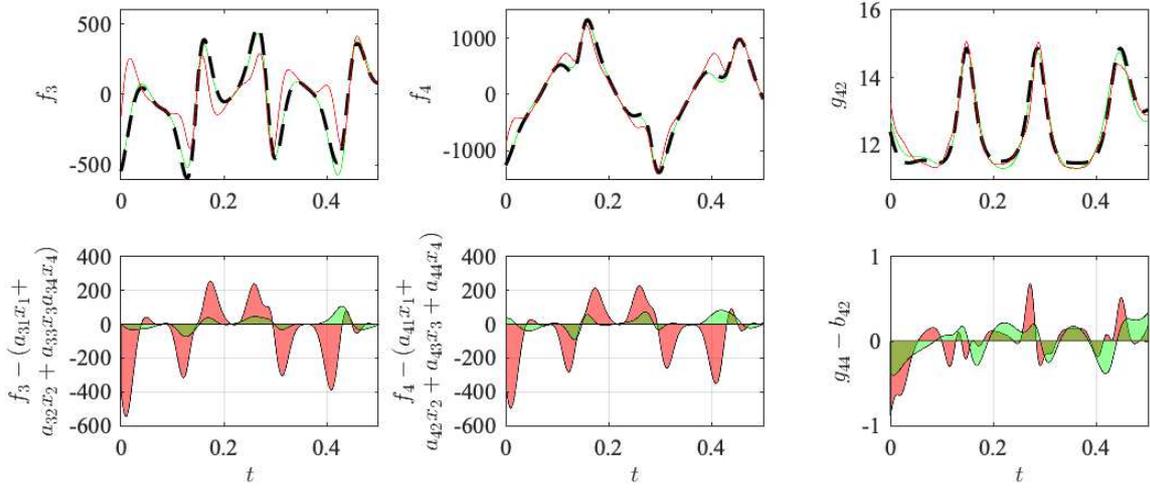}    % The printed column width is 8.4 cm.
		\caption{ \raisebox{+1.5ex}{\colorbox{green!50}{}} Proposed method using 3 scheduling variables. \raisebox{+1.5ex}{\colorbox{red!40}{}} Obtained results by the method of  \citet{HasAbbWer12} using 3 scheduling variables. First row: The actual values for $f_3(x)$, $f_4(x)$, and $g_{42}(x)$ (dashed line)  and their counterparts in the LPV representation. Second row: The error between the actual functions and their counterparts in the LPV representation for the proposed method and the method of \citet{HasAbbWer12}.}
		\label{Fig_comparison}
	\end{center}
	\hrulefill
	\vspace*{4pt}
\end{figure*}

As another example, consider the two-degree of freedom (2-DOF) robot model treated in \cite{HasAbbWer12} which can be modeled as in (\ref{Nonlinear-system}) with the following nonlinear functions: 
 \begin{align*}
&f_1(x)=x_3 , \quad
f_2(x)=x_4  \\
&f_3(x)=\left[ \begin{array}{c}
k_5k_7/V \\ (k_2k_7-k_2k_4)/V \\ -k_3k_4/V \\ k_3^2/(2V) \\-k_3(k_7-k_2)/V \\-k_3k_9/V \\ -(k_6k_7+k_2k_9)/V \\ k_2k_9/V \\k_3k_7/V \\k_3k_9/V
\end{array} \right]^\top
 \left[ \begin{array}{c}
 \sin(x_1) \\\sin(x_2) \\ \sin(x_2)\cos(x_2-x_1) \\x_3^2\sin(2(x_2-x_1)) \\x_3^2\sin(x_2-x_1)\\x_3\cos(x_2-x_1) \\x_3 \\x_4 \\ x_4^2\sin(x_2-x_1) \\x_4\cos(x_2-x_1)
 \end{array}
 \right] 
 \end{align*}
 \[
f_4(x)= \left[ \begin{array}{c}
k_5k_8/V \\-k_3k_5/V \\ (k_4k_8+k_1k_4)/V \\ k_3(k_9+k_6)/V \\ -k_3(k_1+k_8)/V \\ (k_1k_9-k_6k_8)/V \\ k_3k_8/V \\-k_3^2/(2V) \\ -k_1k_9/V \\ -k_3k_9/V 
\end{array}
\right]^\top
\left[ \begin{array}{c}
\sin(x_1) \\ \sin(x_1)\cos(x_2-x_1) \\ \sin(x_2) \\ x_3\cos(x_2-x_1) \\ x_3^2\sin(x_2-x_1) \\ x_3 \\x_3^2\sin(x_2-x_1) \\x_4^2 \sin(2(x_2-x_1))\\x_4 \\x_4\cos(x_2-x_1)
\end{array}
\right]
 \]\[
\left\{ \begin{aligned}
&g_{11}=g_{12}=g_{21}=g_{22}=0 \\
&g_{31}=k_7/V,\quad g_{32}=(-k_3\cos(x_2-x_1)-k_2)/V \\
&g_{41}=(k_8-k_3\cos(x_2-x_1))/V \\
& g_{42}=(k_1+k_3\cos(x_2-x_1))/V
\end{aligned}
\right.
\]
with 
\begin{align*}
V=& k_7k_1+k_2k_8+\\
&\cos(x_2-x_1)(k_7k_3-k_2k_3+k_3k_8-\cos(x_2-x_1)k_3^2)
\end{align*}
and 
\[
k_1=0.0715,~k_2=0.0058,~k_3=0.0114,~k_4=0.3264,\]
\[k_5=0.3957,k_6=0.6253,~k_7=0.0749,~k_8=0.0705,\]\[~k_9=1.1261
\]

For this example, a 1st order polynomial approximation is extracted from the nonlinear functions $f_3$ and $f_4$ and zero-order approximation for $g_{31}$, $g_{32}$, $g_{41}$, and $g_{42}$. Then, different numbers of the scheduling variables are employed to obtain the LPV models. The results are given in Table \ref{Comparison-table}. For the comparison purposes, the related results obtained by the presented method in \cite{HasAbbWer12} which utilizes PCA-based technique of parameter set mapping to obtain a model of complexity low enough is also provided in Table \ref{Comparison-table}. Moreover,  the functions $f_3(x)$, $f_4(x)$, and $g_{42}(x)$ and their counterparts in the LPV modeling are depicted in Fig. \ref{Fig_comparison} for the case of 3 scheduling variables in the first row, and also the errors between the true functions and their counterparts in the LPV representation are depicted in the second row of the same figure. The results show the superiority of our systematic method for the LPV embedding when the same numbers of the scheduling variables are employed.

\begin{table}
	\centering
	\caption{$v_m$ values for the proposed method and the PCA based method}
	\begin{tabular}{ccccc}
		\hline
		Sch. No.        &  3     &  4      & 5      &  6     \\
		\hline
		PCA                      & 0.6960 & 0.8090  & 0.8684 & 0.9210  \\
		Proposed Method & 0.8998 & 0.9810  & 0.9982 & 0.9996 \\
		\hline           
		\label{Comparison-table}
	\end{tabular}
\end{table}

\section{Conclusion}
This paper investigates developing  a systematic and automated LPV embedding for a nonlinear system. A polynomially parameter-dependent model extracted from the nonlinear representation, and to tackle the residuals of the approximation  a set of scheduling variables are introduced. The proposed procedure is composed of polynomial approximation, factorization of the related  residuals, and PCA-based scheduling variable selection, respectively. The obtained LPV model has a scheduling variable vector consisting of the state variables and the introduced scheduling variables; however, imposing some restrictions on the initial approximated model, some of the states of the system may be excluded from being designated as the scheduling variables. One of the main features of the proposed method is that the scheduling variable selection and the LPV embedding conservatism reduction are performed jointly. 
 
\bibliographystyle{ifacconf}
\bibliography{MyCollection1}   

\begin{thebibliography}{23}
\providecommand{\natexlab}[1]{#1}
\providecommand{\url}[1]{\texttt{#1}}
\providecommand{\urlprefix}{URL }
\expandafter\ifx\csname urlstyle\endcsname\relax
  \providecommand{\doi}[1]{doi:\discretionary{}{}{}#1}\else
  \providecommand{\doi}{doi:\discretionary{}{}{}\begingroup
  \urlstyle{rm}\Url}\fi

\bibitem[{Abbas et~al.(2014)Abbas, Toth, Petreczky, Meskin, and
  Mohammadpour}]{AbbTotPet14}
Abbas, H., Toth, R., Petreczky, M., Meskin, N., and Mohammadpour, J. (2014).
\newblock Embedding of nonlinear systems in a linear parameter-varying
  representation.
\newblock In \emph{Proceedings of the 19th IFAC World Congress of the
  International Federation of Automatic Control, (IFAC'14), 24-29 August 2014,
  Cape Town, South Africa}, 6907--6913.

\bibitem[{Apkarian et~al.(1995)Apkarian, Gahinet, and Becker}]{ApkGahBec95}
Apkarian, P., Gahinet, P., and Becker, G. (1995).
\newblock {Self-scheduled {$H_\infty$} control of linear parameter-varying
  systems: a design example}.
\newblock \emph{Automatica}, 31(9), 1251--1261.

\bibitem[{Bachnas et~al.(2014)Bachnas, T{\'{o}}th, Ludlage, and
  Mesbah}]{BacTotLud14}
Bachnas, A.A., T{\'{o}}th, R., Ludlage, J.H.A., and Mesbah, A. (2014).
\newblock {A review on data-driven linear parameter-varying modeling
  approaches: A high-purity distillation column case study}.
\newblock \emph{Journal of Process Control}, 24(4), 272--285.

\bibitem[{Caigny et~al.(2011)Caigny, Camino, and Swevers}]{CaiCamSwe11}
Caigny, J.D., Camino, J.F., and Swevers, J. (2011).
\newblock {Interpolation-Based Modeling of {MIMO} {LPV} Systems}.
\newblock \emph{IEEE Transactions on Control Systems Technology}, 19(1),
  46--63.

\bibitem[{Cand{\`{e}}s et~al.(2008)Cand{\`{e}}s, Wakin, and Boyd}]{CanWakBoy08}
Cand{\`{e}}s, E.J., Wakin, M.B., and Boyd, S.P. (2008).
\newblock {Enhancing Sparsity by Reweighted $l_1$ Minimization}.
\newblock \emph{Journal of Fourier Analysis and Applications}, 14(5), 877--905.

\bibitem[{Chisci et~al.(2003)Chisci, Falugi, and Zappa}]{ChiFalZap03}
Chisci, L., Falugi, P., and Zappa, G. (2003).
\newblock Gain-scheduling mpc of nonlinear systems.
\newblock \emph{International Journal of Robust and Nonlinear Control},
  13(3‐4), 295--308.

\bibitem[{Daafouz et~al.(2008)Daafouz, Bernussou, and Geromel}]{DaaBerGer08}
Daafouz, J., Bernussou, J., and Geromel, J.C. (2008).
\newblock {On Inexact {LPV} Control Design of Continuous-Time Polytopic
  Systems}.
\newblock \emph{IEEE Transactions on Automatic Control}, 53(7), 1674--1678.

\bibitem[{{De Caigny} et~al.(2012){De Caigny}, Camino, Oliveira, Peres, and
  Swevers}]{CaiCamOli12}
{De Caigny}, J., Camino, J.F., Oliveira, R.C.L.F., Peres, P.L.D., and Swevers,
  J. (2012).
\newblock {Gain-scheduled dynamic output feedback control for discrete-time
  {LPV} systems}.
\newblock \emph{International Journal of Robust and Nonlinear Control}, 22(5),
  535--558.

\bibitem[{Hashemi et~al.(2012)Hashemi, Abbas, and Werner}]{HasAbbWer12}
Hashemi, S.M., Abbas, H.S., and Werner, H. (2012).
\newblock Low-complexity linear parameter-varying modeling and control of a
  robotic manipulator.
\newblock \emph{Control Engineering Practice}, 20(3), 248 -- 257.

\bibitem[{Henson and Seborg(1998)}]{HenSeb98}
Henson, M. and Seborg, D. (1998).
\newblock \emph{{Nonlinear process control}}.
\newblock Prentice Hall, Prentice Hall, Englewood Cliffs, NJ, USA.

\bibitem[{{Hoffmann} and {Werner}(2015)}]{HofWer15}
{Hoffmann}, C. and {Werner}, H. (2015).
\newblock A survey of linear parameter-varying control applications validated
  by experiments or high-fidelity simulations.
\newblock \emph{IEEE Transactions on Control Systems Technology}, 23(2),
  416--433.

\bibitem[{{Kwiatkowski} et~al.(2006){Kwiatkowski}, {Boll}, and
  {Werner}}]{KwiBolWer06}
{Kwiatkowski}, A., {Boll}, M., and {Werner}, H. (2006).
\newblock Automated generation and assessment of affine lpv models.
\newblock In \emph{Proceedings of the 45th IEEE Conference on Decision and
  Control}, 6690--6695.

\bibitem[{{Kwiatkowski} and {Werner}(2008)}]{KwiWer08}
{Kwiatkowski}, A. and {Werner}, H. (2008).
\newblock {PCA}-based parameter set mappings for {LPV} models with fewer
  parameters and less overbounding.
\newblock \emph{IEEE Transactions on Control Systems Technology}, 16(4),
  781--788.

\bibitem[{Nijmeijer and van~der Schaft(1990)}]{NijSch90}
Nijmeijer, H. and van~der Schaft, A. (1990).
\newblock \emph{{Nonlinear dynamical control systems}}.
\newblock Springer Verlag, New York, Berlin, Heidelberg.

\bibitem[{Oliveira and Peres(2007)}]{OliPer07}
Oliveira, R.C.L.F. and Peres, P.L.D. (2007).
\newblock {Parameter-Dependent {LMIs} in Robust Analysis: Characterization of
  Homogeneous Polynomially Parameter-Dependent Solutions Via {LMI}
  Relaxations}.
\newblock \emph{IEEE Transactions on Automatic Control}, 52(7), 1334--1340.

\bibitem[{{Petersson} and {Löfberg}(2009)}]{PetLof09}
{Petersson}, D. and {Löfberg}, J. (2009).
\newblock Optimization based {LPV}-approximation of multi-model systems.
\newblock In \emph{2009 European Control Conference (ECC)}, 3172--3177.

\bibitem[{{Sadeghzadeh}(2019)}]{SadIET19}
{Sadeghzadeh}, A. (2019).
\newblock {LMI} relaxations for robust gain-scheduled control of uncertain
  linear parameter varying systems.
\newblock \emph{IET Control Theory Applications}, 13(4), 486--495.

\bibitem[{Sadeghzadeh(2018{\natexlab{a}})}]{SadIJRNC18}
Sadeghzadeh, A. (2018{\natexlab{a}}).
\newblock Gain-scheduled continuous-time control using polytope-bounded inexact
  scheduling parameters.
\newblock \emph{International Journal of Robust and Nonlinear Control}, 28(17),
  5557--5574.

\bibitem[{Sadeghzadeh(2018{\natexlab{b}})}]{SadSCL18}
Sadeghzadeh, A. (2018{\natexlab{b}}).
\newblock On exploiting inexact scheduling parameters for gain-scheduled
  control of linear parameter-varying discrete-time systems.
\newblock \emph{Systems \& Control Letters}, 117, 1 -- 10.

\bibitem[{Sato and Peaucelle(2013)}]{SatPea13}
Sato, M. and Peaucelle, D. (2013).
\newblock {Gain-scheduled output-feedback controllers using inexact scheduling
  parameters for continuous-time {LPV} systems}.
\newblock \emph{Automatica}, 49(4), 1019--1025.

\bibitem[{Schoukens and Toth(2018)}]{SchTot18}
Schoukens, M. and Toth, R. (2018).
\newblock Linear parameter varying representation of a class of mimo nonlinear
  systems.
\newblock \emph{IFAC-PapersOnLine}, 51(26), 94 -- 99.
\newblock 2nd IFAC Workshop on Linear Parameter Varying Systems LPVS 2018.

\bibitem[{{Sznaier} and {Mazzaro}(2003)}]{CznMaz03}
{Sznaier}, M. and {Mazzaro}, M.C. (2003).
\newblock An {LMI} approach to control-oriented identification and model (in)
  validation of {LPV} systems.
\newblock \emph{IEEE Transactions on Automatic Control}, 48(9), 1619--1624.

\bibitem[{Toth(2010)}]{Tot10}
Toth, R. (2010).
\newblock \emph{Modeling and identification of linear parameter-varying
  systems}.
\newblock Lecture notes in control and information sciences. Springer, Germany.
\newblock \doi{10.1007/978-3-642-13812-6}.

\end{thebibliography}
\end{document}